\documentclass[journal=jacsat,manuscript=article]{achemso}
\usepackage[version=3]{mhchem} 

\newcommand*{\rmb}[1]{\boldsymbol{\rm #1}}

\author{Kezheng Zhu}
\author{Erich A. M\"uller}
\email{e.muller@imperial.ac.uk}
\affiliation[Imperial College London]{Department of Chemical Engineering, Imperial College London, U.K.}

\title{Generating a Machine-learned Equation of State for Fluid Properties}

\abbreviations{}

\begin{document}
\begin{tocentry}
\includegraphics[scale=1]{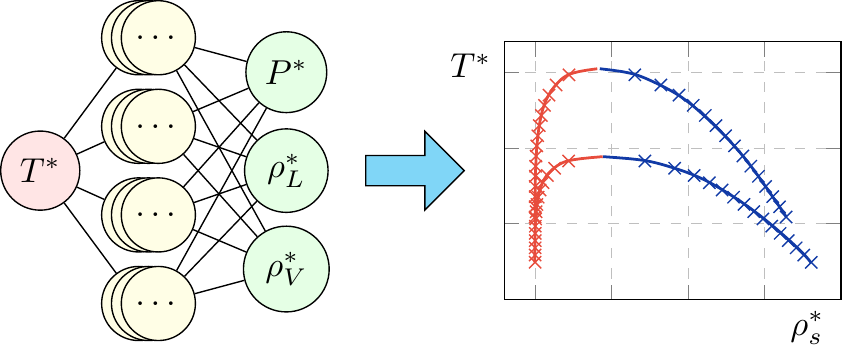}
\end{tocentry}
\begin{abstract}
    Equations of State (EoS) for fluids have been a staple of engineering design and practice for over a century. Available EoS are based on the fitting of a closed-form analytical expression to suitable experimental data. The underlying mathematical structure and the underlying physical model significantly restrain the applicability and accuracy of the resulting EoS. This contribution explores the issues surrounding the substitution of analytical EoS for machine-learned models, in particular, we describe, as a proof of concept, the effectiveness of a machine-learned model to replicate statistical associating fluid theory (SAFT-VR-Mie) EoS for pure fluids. By utilizing Artificial Neural Network and Gaussian Process Regression, predictions of thermodynamic properties such as critical pressure and temperature, vapor pressures and densities of pure model fluids are performed based on molecular descriptors. To quantify the effectiveness of the Machine Learning techniques, a large data set is constructed using the comparisons between the Machine-Learned EoS and the surrogate data set suggest that the proposed approach shows promise as a viable technique for the correlation, extrapolation and prediction of thermophysical properties of fluids.
\end{abstract}

\section{Introduction}

The virtual issue where this contribution is included and its predecessor\cite{schneider2018machine} are devoted to showcasing examples where machine learning (ML) has been employed to contribute to the physical sciences. The reader is referred to the indexes of these issues for examples and reviews of the most recent advances in this rapidly evolving topical field. Inspection of this, and other literature, suggests that while the pharma and biomedical research groups have, in growing numbers, been successfully employing ML to, for example, explore candidate drug molecules\cite{pereira2016boosting,subramanian2016computational}, or the design of new bioactive ones\cite{ash2017characterizing}, the application of ML in the physical and chemical sciences has been much less prevalent\cite{sumpter1994theory,schneider2018machineB}. In particular, a much slower progress has been evidenced in the field of engineering, with salient counterexamples in process system modelling\cite{venkatasubramanian1989neural} and the prediction of properties of inorganic materials\cite{sumpter2015bridge,ward2016general,spellings2018machine}. Given this general scenario, this manuscript deals with the premise that the currently available hardware and software can allow the wide-spread implementation of ML to the prediction and correlation of thermophysical properties of fluids.

Physical properties of fluids can, and have, been measured experimentally since the dawn of modern science. Densities, vapor pressures, critical point data, viscosities, solubilities, surface tensions, thermal conductivities, etc. can all be routinely measured and are collated in existing open-access and closed-source databanks\cite{chemspider,dippr,nist,dechema}. Pure compound data is presently available for several thousand to tens of thousands chemical compounds of interest (depending on the databank), which pales against the 84 million chemical structures positively identified to date\cite{chemspider} and the 166.4 billion possible molecules of up to 17 atoms of C, N, O, S and halogens collected in GBD-17\cite{ruddigkeit2012enumeration}. When one considers mixtures, a much smaller (relative) number of systems have been explored. The slow pace (and cost) of experimental acquisition of data is at odds with the massive phase space of pure and mixture systems of interest. Databases themselves grow at a slow pace, to the order of 1 M data points per year. Entire scientific journals are devoted to this pursuit and the progress is very incremental, at the most.

The current engineering folklore for dealing with this lack of data is to use empirical correlations and/or fitted simplified theories to essentially interpolate and/or extend the results. Historians of Thermodynamics have given detailed account of the quest of scientists to rationalize the interrelation between the thermophysical properties of fluids\cite{muller2007history,rowlinson2005cohesion}. On the other hand, the prediction of properties by quantum mechanical calculations are restricted to small, relatively simple molecules and the use of molecular simulation using empirical force-fields also has limits in terms of the speed of calculation and the estimation of the expected accuracy\cite{ungerer2007molecular,nieto2015general}. Given this scenario, there is still a pressing need for the correlation, and possible generalization, of available experimental data. 

The correlation of volumetric properties of fluids has it most renowned exemplary in the 1910 physics Nobel prize awarded to J.D. van der Waals, for the recognition that analytical (mathematical) models were capable of modelling vapor-liquid equilibria. Surprisingly, a century later, these simple expressions, better known today as cubic Equations of State (EoS) are a staple of engineering design. Expectedly, the field has advanced and extremely refined EoS, are now available\cite{kontogeorgis2009thermodynamic}. Current propositions are based on our interpretation of the molecular nature of matter. i.e. theories start with an expected simplified molecular model and develop corresponding approximations in order to arrive to a closed form expression. The resulting model is further commonly employed as correlation tool, with parameters fit to experimental properties. There are several restricting limitations to this approach as current EoS will commonly have limits with respect to the range of chemical interactions (e.g. length and shape of molecules, polarity, presence of ions) and the availability of pure fluid and mixture parameters. 

The quest for employing ML in Thermodynamics as a correlation tool is not new, but improvements in the available hardware and software have produced a resurgence in the field\cite{venkatasubramanian2019promise}. Most of the effort has been focused on employing artificial neural networks (ANN), mostly due to the ease in which they can be deployed (see for example ref.\cite{joss2019machine} for classroom examples of creating boiling point correlations with ANN). A review of some of the first applications of ANN to the chemical sciences and engineering was provided by Himmelblau\cite{himmelblau2008accounts}. Further to that, ANNs have been used to selectively correlate a limited number of thermophysical properties of restricted families of compounds, for example alkanes\cite{pirdashti2020thermophysical,santak2019predicting}, ionic liquids\cite{golzar2014prediction}, refrigerants\cite{csencan2011prediction,azari2013prediction}, components of biofuels\cite{saldana2012prediction}, gases\cite{coccia2019determination,bouzidi2007viscosity}. Critical properties\cite{hall1996boiling}, Interfacial properties\cite{vasseghian2019modeling} and partition coefficients\cite{huuskonen2000neural,lowe2011comparative} have all been individually explored. The reader is referred to an excellent recent review by Forte et al\cite{forte2019digitalization} and references therein for a modern account. 

In most of the aforementioned examples, the onus has been on the correlation of a selected and well-chosen sub-set of properties of a well-defined family of chemical compounds. As such, the versatility and universality of common engineering methods (e.g. EoS) has not been matched. This contribution aims at taking a step in this direction, exploring the question of whether current ML approaches can be employed to substitute analytical EoS. In a seminal contribution, Arce et al.\cite{arce2018thermodynamic} compared the quality of fit of common EoS for binary systems involving solvents and biodiesel components at supercritical conditions to that of an ANN, finding a good agreement between both the EoS, the ANN and experimental data. The work of Arce et al. suggests that an ANN could be built to substitute an analytical EoS. But further questions arise, is the error associated with the correlation acceptable? More importantly, are ANN the optimal choice of a ML algorithm for this purpose? A recent contribution\cite{cendagorta2020comparison} pointed towards ANNs as being the most robust for this purpose, but lauded the benefits of kernel-based methods, leaving the matter unsolved.

To assess the above-posed questions, in a most general (and ideal scenario), one would start with experimental data and use part of the database as a training set to fit a ML model. However, experimental data is both noisy and very poorly distributed; a larger abundance of data points are at experimentally “easy” conditions, which may or may not be the most appropriate to describe the whole phase space. To circumvent this, we chose to generate pseudo-experimental data from a current state-of-the-art analytical EoS, the Statistical Associating Fluid Theory – variable range Mie equation\cite{Lafitte2013} (SAFT-VR-Mie). This procedure, developed in this contribution, allows us to a) test the capacity of the ML algorithms to correlate the data, b) To generate a large amount of pseudo-data covering all regions of the relevant phase space, and most importantly c) allows for a well-defined characterization of the error associated with the ML fitting.

\subsection{SAFT Equation of State}

The SAFT-VR-Mie framework is a coarse-grained model wherein molecules are depicted as collections of tangent spherical segments, colloquially termed as beads. The interactions acting between two identical segments follows the Mie potential (a generalized Lennard-Jones potential), which is specified by four force field parameters ($\sigma$, $\varepsilon$, $\lambda_r $, $\lambda_a$). The distinguishing aspect of the SAFT force field model is that there is a direct correspondence between the underlying force field and the analytical EoS, i.e. the molecular simulations and the EoS produce the same outputs for a given set of molecular parameters. In terms of what concerns us here, the advantage of the SAFT formulation is that molecules are described through a set of descriptors that relate directly to molecular properties. A wide range of real substances has been shown to be modelled effectively as chains of coarse-grained beads\cite{Avendano2011,Papaioannou2014}, from which fluid behaviour and thermodynamic properties can be determined. We stress from the onset that the focus of this contribution is not to correlate experimental data and/or to produce the ultimate EoS, but to assess the extent to which a ML model can replace an analytical EoS in a well defined context. The underlying force field is based on the Mie potential \cite{Mie1903,Gruneisen1912}:
\begin{equation}\label{eq:miepot}
\mathcal{U}_\text{Mie}(r_{ij}) = C_\text{Mie} \varepsilon \left[ \left(\frac{\sigma}{r_{ij}}\right)^{\lambda_r} -  \left(\frac{\sigma}{r_{ij}}\right)^{\lambda_a} \right]\\
\end{equation}
where $r$ is the inter-segment distance, $\varepsilon$ is the potential depth, $\sigma$ is the segment diameter (or position at which the potential is zero), and $\lambda_r$ and $\lambda_a$ are the repulsive and attractive exponents which characterize the pair-wise energy. The constant $C_\text{Mie}$ is defined as:
\begin{equation}
C_\text{Mie}  = \left( \frac{\lambda_r}{\lambda_r-\lambda_a} \right) \left( \frac{\lambda_r}{\lambda_a} \right)^{\frac{\lambda_a}{ \lambda_r-\lambda_a}} \label{eq:cmie}
\end{equation}
Tying together the Mie potential and thermodynamic properties is the SAFT-VR-Mie equation of state\cite{Lafitte2013}, a closed analytical form of the Helmholtz free energy for chains of spherical segments interacting through the Mie potential. The SAFT-VR-Mie is a special case of the SAFT-$\gamma$ model\cite{Avendano2011b} where molecules are represented by a collection of homonuclear beads with no distinction between functional groups in a molecule. The use of the SAFT-$\gamma$ model implies a higher level of complexity which is irrelevant to the objectives of this study. The reader is referred to \citeauthor{Lafitte2013}\cite{Lafitte2013} for the explicit expressions in the SAFT-VR-Mie equation of state.

In a typical use of Mie force field parameters ($\sigma, \varepsilon, \lambda_r, \lambda_a$) to represent a fluid, we look to the SAFT-VR-Mie EoS to estimate the parameters that provide the best representation of available macroscopic experimental data. This procedure has been performed for thousands of fluids, and the parameters are available in the web tool named Bottled SAFT.\cite{Ervik2016}. As the EoS explicitly calculates Helmholtz free energy, macroscopic thermodynamic properties such as the vapor pressure $P_v$ and saturated liquid density $\rho_L$ and equilibria vapor-liquid conditions can be derived directly by standard thermodynamic relationships. It should be noted that other second-derivative properties such as Joule-Thomson coefficient, heat capacity and speed of sound can also be extracted from the EoS. 

By using the EoS to produce fluid phase equilibria data for a wide range of descriptors corresponding to hypothetical ``Mie fluids", a vast amount of data can be generated for the use of training a ML model. The following sections will very briefly describe two algorithms which were employed in this work: artificial neural network (ANN) and Gaussian process regression (GPR).

\subsection{Artificial Neural Network (ANN)}
ANN is a computational model partly inspired by biological neural systems, where interconnected neurons create a network of mathematical operations to model multi-dimensional problems\cite{sha2007use,goodfellow2016deep}. The most common type of ANN is a feed-forward neural network, where neurons are ordered in interconnected layers: an input layer, hidden layer(s) and an output layer. Each neuron represents a single unit of computation, where it receives a set of inputs for which a weight is assigned based on relative importance. Taking the weighted sum of its inputs, the neuron then generates an output by adding a bias. Mathematically, output $\rmb{y}$ of a single layer with $m$ neurons given an input vector $\rmb{x}$ of size $n$ can be expressed as:
\begin{equation}
\rmb{y} = \rmb{W}\rmb{x} + \rmb{b}
= \begin{bmatrix}
w_{1,1} & w_{1,2} & \dots & w_{1,n} \\
\vdots & \vdots & \ddots & \vdots \\
w_{m,1} & w_{m,2} & \dots & w_{m,n}
\end{bmatrix}
\begin{bmatrix}
x_{1}\\
\vdots \\
x_{n}
\end{bmatrix}
+
\begin{bmatrix}
b_{1}\\
\vdots \\
b_{m}
\end{bmatrix}
\end{equation}
where $\rmb{w}$ is the weight matrix ($m\times n$) for each input element and each neuron, $\rmb{b}$ is the bias vector corresponding to the layer.

The linear output of each neuron will also be passed through an activation function to introduce non-linearity into an ANN model, with common functions such as the sigmoid function, hyperbolic tangent function and exponential linear unit (ELU). The continuity of derivatives for the ANN model is an important criterion when considering the choice of activation functions, for which the Tanh and ELU functions are used:
\begin{gather}
\intertext{The Tanh or hyperbolic tangent activation function transforms values of x to a range between -1 and 1,}
f(x) = \tanh (x) = \dfrac{1 - e^{-2x}}{1 + e^{-2x}}\\
\intertext{while the ELU or Exponential linear unit changes the value of x to an exponential decay.}
f(x) = 
\begin{cases}
    x & \qquad x > 0\\
    \alpha \left( e^{x} - 1 \right) & \qquad x < 0
\end{cases}
\end{gather}
Mathematically, a $N$-layered neural network can be written as:
\begin{equation}\label{eq:ann}
\rmb{y} = f_N \left( \rmb{W}_N \dots \left( \rmb{W}_2 f_1( \rmb{W}_1 \rmb{x} + \rmb{b}_1 ) + \rmb{b}_2 \right) \dots + \rmb{b}_N \right)
\end{equation}
To train the ANN, supervised learning is employed. By providing the input data (or features) and output data (or labels), the ANN is trained by evaluating the inputs, comparing the outputs and adjusting the parameters. The neural network is initialized with randomly assigned weights and biases, while employing an optimizer such as stochastic gradient descent (SGD) or the Adam optimizer\cite{Kingma2014} to minimize the loss function defined by the mean squared error (MSE) between the predicted outputs and the training data.

Aside from a common feed-forward neural network, there are multiple types of ANN developed for different purposes. Convolutional layers in neural network (CNN) corresponds to inputs arranged in a 2 dimensional matrix, employs filter (or kernel) in hidden layers to recognize edges for image recognition\cite{simard2003best}. Recurrent neural network (RNN) works memory storage in its neurons such that past iterations has an effect on current iteration, primarily used for time series or problems involving unknown number of features\cite{chung2014empirical}. Different network structures exist outside of feed-forward neural networks, with architectures such as bridged multi-layer (BMLP) and fully connected cascade (FCC)\cite{wilamowski2009neural} where non-adjacent layers are connected such that there are more adjustable parameters for lesser neurons. With the wide variety of neuron cells, activation functions and possible network architectures, ANN has a key advantage of flexibility in implementation amongst machine learning algorithms.

In this work, we used a multilayer perceptron ANN, referring to a feed-forward neural network with multiple hidden layers, along with the SGD optimizer. The SGD optimizer is common staple of optimization in deep learning, where gradient descent towards the objective function (minimizing loss function) is done with an estimate of the gradient, considering a random subset of data to reduce computational burden. Although the Adam optimizer is widely recommended as the default optimizer for deep learning\cite{ruder2016overview}, the adaptive nature of the optimization algorithm suits problems with sparse gradients and/or noisy problems, which is not required here.

\subsection{Gaussian Process Regression (GPR)}

A Gaussian process (GP) is defined as a collection of Gaussian distributed random variables $f$, for which a function can be represented as an infinitely long vector of such random variables $[f_1, f_2, \dots]$ \cite{williams2006gaussian}. Each distribution or point exhibits a corresponding mean $\mu$ and a covariance vector $\Sigma$, while the function itself is specified by a mean function $m(x_i)$ (the average of the distribution for all the random variables) and a covariance function $k(x_i, x_j)$ (kernel) \cite{williams2006gaussian}. 

A GP model ``learns" by ``observing" data points and deriving the posterior distribution from the prior by determining the conditional probability through the covariance matrix. In simple terms, by treating the functions as random variables $f_i$, any two variable is correlated through the covariance function (or kernel). By observing a single point $f_a$, or a set of values, corresponding to a point in the phase space, any other point in the phase space can be associated with a new observed value using the prior distribution and a Gaussian likelihood function. The result is a posterior distribution, where the mean, variance and covariance matrix can be derived from the prior. Repeating the process for the whole data set, the GP model ``learns" as each data point is ``observed". 

The covariance function, or kernel, in GP regression (GPR) determines the shape and smoothness of the mean function. In this work, the use of the Radial-basis function (RBF) and the Mat\'ern kernel is explored.

The radial basis function (RBF) is also known as the squared exponential covariance function, can be expressed as:
\begin{equation}\label{eq:rbf}
K \left( x_i, x_j \right) = \sigma^2 \exp \left( -\dfrac{|| x_i - x_j ||^2}{2l^2} \right)
\end{equation}
where $|| x_i - x_j ||^2$ is the Euclidean distance between the two data points $x_i$ and $x_j$, $l$ is a trainable length scale parameter and $\sigma^2$ is the variance. Taking $d$ as the Euclidean distance, the Mat\'ern kernel\cite{handcock1993bayesian} is expressed as:
\begin{equation}
K \left( x_i, x_j \right) = \sigma^2 \dfrac{2^{1-\nu}}{\Gamma (\nu)} \left( \sqrt{2\nu} \dfrac{d}{l} \right)^\nu K_\nu \left( \sqrt{2\nu} \dfrac{d}{l} \right)
\end{equation}
where $\Gamma$ is the gamma function, $K_\nu$ is the modified Bessel function, and $\nu$ is an additional parameter for which larger $\nu$ indicates a smoother covariance function. The Mat\'ern function ($C_\nu$) converges to RBF as $\nu$ approaches infinity. Most commonly, $\nu = 1/2$, $3/2$ and $5/2$ are used, corresponding to the Matern12, Matern32 and Matern52 kernel:
\begin{align}
C_{\nu=1/2} &= \sigma^2 \exp \left( -\dfrac{d}{l} \right) \\
C_{\nu=3/2} &= \sigma^2 \left(1 + \dfrac{\sqrt{3}d}{l} \right) \exp \left( -\dfrac{\sqrt{3}d}{l} \right) \\
C_{\nu=5/2} &= \sigma^2 \left(1 + \dfrac{\sqrt{5}d}{l} + \dfrac{5d^2}{3l^2} \right) \exp \left( -\dfrac{\sqrt{5}d}{l} \right) 
\end{align}
It is important to note that the Mat\'ern kernel is differentiable to $\lceil \nu \rceil - 1$: the 1/2 kernel is not differentiable, 3/2 is once differentiable and 5/2 is twice differentiable. Naturally, the RBF is infinitely differentiable. The reader is referred to \citeauthor{williams2006gaussian}\cite{williams2006gaussian} for detailed explanation of GPR and the kernel functions.

The key advantage is using GPR is its ability to generate accurate models with less but noise-free data, or data with known error margin. This is advantageous for models where the user knows the inherent smoothness and/or function shape of the model. With the ability to calculate variance of the output function, GPR also has a built-in error indicator, which can be advantageous for researchers to determine the next data point to conduct experiment or simulation to improve the model. The drawback with GPR is the computation complexity: due to the inversion of the covariance matrix, the computation time scales in $O(n^3)$ with respect to data size of $n$.\cite{krishnamoorthy2013matrix}

\section{Methodology}

In this study, we examine the ability of both non-linear ANN and GPR to correlate thermodynamic properties in the way a traditional EoS would. Since the ML algorithm relies on the processing of large amounts of data, the direct use of experimental data might turn out to produce inconclusive results, as the typical data points will be biased in quantity and quality towards the experimentally ``easier" state points (e.g. room temperature and/or pressure). By generating a larger database of pseudo-experimental data employing the SAFT EoS, we remove these limitations and biases and gauge both the ability of ML models to recognize and correlate the data, and are able, by direct comparison to the EoS, to quantify the associated error. Within the SAFT-VR-Mie framework, the parameters which describe a molecule are $m_s$, $\sigma$, $\varepsilon$, $\lambda_r$ and $\lambda_a$. In this particular scenario, the conformality of the Mie potential\cite{ramrattan2015corresponding} helps reduce the dimensionality of the problem by setting $\lambda_a$ to 6 and scaling properties with respect to the length ($\sigma$) and energy ($\varepsilon$) scale. The result is that a given molecule can be uniquely characterized by the number of spheres $m_s$ and the repulsive exponent $\lambda_r$\cite{Ervik2016,mejia2014force}. In particular, we look at the dimensionless form of temperature $T$, pressure $P$, and segment density $\rho$.
\begin{align}
T^* &= \dfrac{kT}{\varepsilon} \\
P^* &= \dfrac{P\sigma^3}{\varepsilon} \\
\rho^* &= \rho \sigma^3
\end{align}
Although we fix the attractive parameter to $\lambda_a = 6$, the combination of repulsive and attractive exponents can be replaced with a unique parameter $\alpha$, where different exponent pairs ($\lambda_r$,$\lambda_a$) giving the same value of $\alpha$ parameters provide the same dimensionless vapor-liquid equilibrium (VLE) behaviour\cite{ramrattan2015corresponding}. The $\alpha$ parameter is given by:
\begin{equation}
\alpha = C_\text{Mie} \left[ \left(\dfrac{1}{\lambda_a - 3}\right) - \left(\dfrac{1}{\lambda_r - 3}\right) \right]
\end{equation}
hence $\alpha$ and $\lambda_r$ may be used indistinctively to characterize the shape of the potential.

To compare the effectiveness of ANN and GPR for different levels of complexity, we attempt to replicate the equation of state for critical properties, VLE properties and supercritical density. Each set of properties corresponds to a problem with different complexity, being a 2-dimensional problem (2 input features) for critical properties, 3-dimensional problem for VLE properties and 4-dimensional problem for supercritical (or one phase) density.

For ANN, the choice of the number of nodes and layers is linked to the complexity of the problem. The approach used here is to gradually increase the number of layers and nodes just until the accuracy of the predicted model plateaus to prevent over-fitting. For more than 1 hidden layer, subsequent layers have lesser nodes in order to compress information approaching the output\cite{huang2006universal}. As we are employing ANN for a regression problem, the activation functions used should be a non-linear continuous function. The tanh function is chosen as it is infinitely differentiable and has continuous derivatives (the second derivative of an ELU function is discontinuous), a property that is important in the applications of thermodynamics. The overall data sets are also split into a training and validation set, with a ratio of 0.8 training data vs 0.2 validation data which the model will not see and will be validated against. The implementation of the ANN model is done using PyTorch\cite{pytorch2019}.

For GPR, a linear combination of the Radial Basis Function and a linear kernel for each dimension is used to ensure that the produced model is infinitely differentiable. Due to the significant increase in computational time for each additional data point and the low marginal benefit of increasing dataset beyond a certain limit, we find that keeping the dataset between 500 to 1000 points results in sufficiently accurate models for analysis.

\subsection{Data Transformations}

\begin{figure}[tbp!]
	\centering
	\includegraphics[scale=0.95]{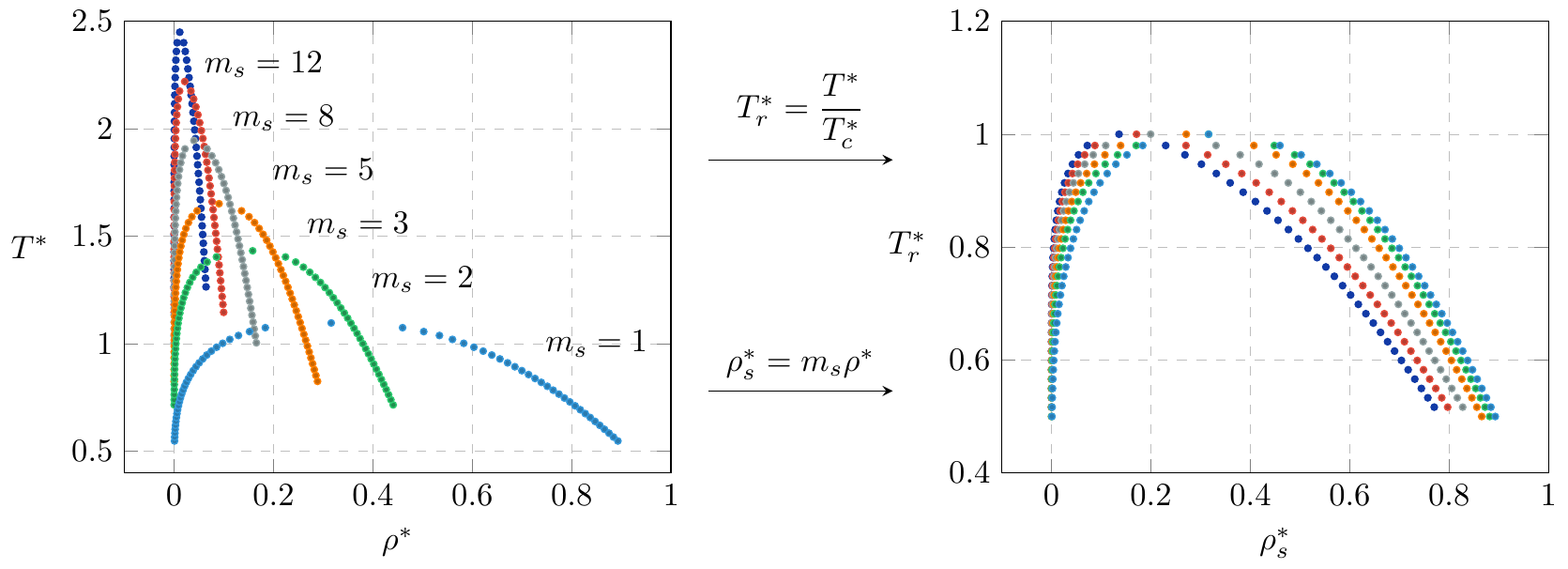}
	\caption{Plot of selected generated data for reduced temperature, $T^*$, (with respect to energy scale) against saturated liquid and vapor density, $\rho^*$, for molecules with the same repulsive exponent $\lambda_r = 16$ but different number of segments $m_s$ (left). Regularization is accomplished by scaling the temperature with respect to the critical point, $T^*_c$, and expressing the molecular density in terms of the segment density, $\rho^*_s$. In this way, direct and known relationships (such as that between density and $m_s$) are removed, allowing for the model to regress solely the subtle changes in response to different molecular descriptors (right).} 
	\label{fig:trans}
\end{figure}%
\begin{figure}[tbp!]
	\centering
	\includegraphics[scale=0.95]{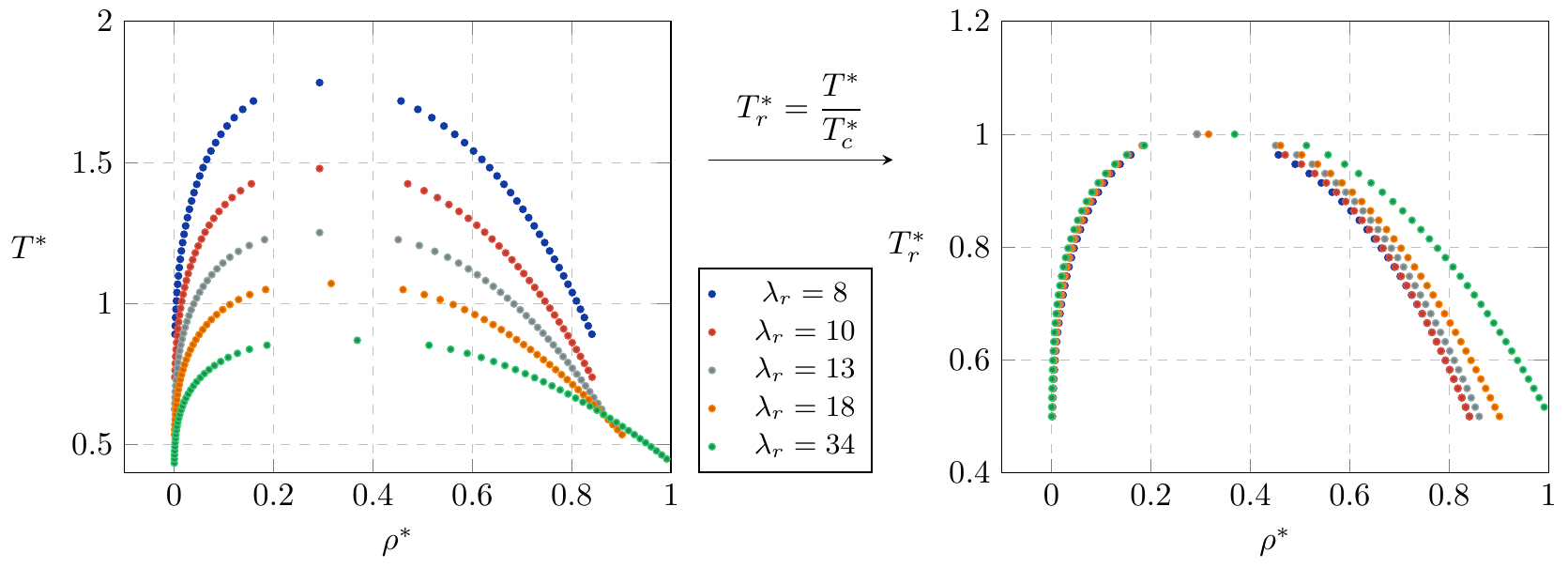}
	\caption{Plot of selected generated data for reduced temperature, $T^*$, (with respect to energy scale) against saturated liquid and vapor density, $\rho^*$, for molecules with the same number of segments $m_s = 1$ but different repulsive exponent $\lambda_r$ (left). Similar to Figure \ref{fig:trans}, scaling temperature with respect to critical temperature results in a more concise data set for modelling (right).}
	\label{fig:trans2}
\end{figure}%
Although, in principle, ML algorithms seem capable of deducing the relationships amongst explicit and/or hidden properties, a more robust prediction can be obtained if one can incorporate some known information in the model. Furthermore, regularization (understood here as an appropriate scaling of the data) is a key trick of the trade in ML, as it reduces bias in the correlation of data. Prior knowledge of simple thermodynamic relationships helps in this respect. For example, the boiling point $T^*$ and vapor pressure $P^*_v$ for the VLE envelope converges at the critical point ($T^*_c, P^*_c$), hence scaling the temperature inputs and vapor pressure outputs with the critical temperature and pressure of a molecule respectively reduces the impact of differing critical point for molecules while scaling the absolute values to the range [0,1]. Similarly, the saturated liquid and vapor density expressed in the \textbf{molecular} density $\rho^*$ differs greatly with respect to $m_s$, while the \textbf{segment} density $\rho^*_s = m_s \rho^*$ for all molecules varies in a tighter range. 

Figure \ref{fig:trans} illustrates the variation of the VLE phase envelope for the same repulsive exponent $\lambda_r$ but different number of segments $m_s$, and the effect of data transformation based on existing knowledge. Similarly, Figure \ref{fig:trans2} exemplifies the effect of scaling with respect to critical temperature when considering different repulsive exponents for the same $m_s$. By introducing prior scientific knowledge of the system behaviour into transforming the data, the data-driven model can be trained more effectively to the desired correlations we wish to observe. 

\subsection{Critical Properties}

The critical point on the phase diagram of any fluid corresponds to a temperature, $T^*_c$, and pressure, $P^*_c$, where the vapor-liquid boundary terminates. The critical point has zero degrees of freedom as for any pure substance, the critical point can only exist at one temperature and one pressure. In order to predict critical properties from the molecular descriptors, the problem could be defined as:
\begin{equation}
\mathcal{F}_1 : (m_s, \alpha) \to (T_c^*, P_c^*)
\end{equation}
where $\mathcal{F}_i$ refers to the specific relation we seek to discover.

Using the SAFT-VR Mie equation of state, the critical properties can be obtained by using a solving algorithm for the stated condition above. By sampling different repulsive exponents $\lambda_r$ (directly related to $\alpha$, as $\lambda_a = 6$) at a typical range of 8 - 34 and number of segments $m_s$ at 1 to 20, 540 data points are generated in order to train a machine learning model. 

\subsection{VLE Properties}

For a pure fluid in vapor-liquid equilibrium, the two coexisting phases must satisfy the mechanical and diffusive equilibrium conditions. Provided that the pressure of both phases is equal, $P_L = P_V$, and the Gibbs free energy (or chemical potential) is also equal, $G_L = G_V$, one is left with a single degree of freedom. This allows us to formulate a three-dimensional problem with respect to temperature $T$ to predict VLE properties vapor pressure $P_v$, saturated liquid density $\rho_L$ and saturated vapor density $\rho_V$.
\begin{equation}
\mathcal{F}_2 : (m_s, \alpha, T_r^*) \to (P_{v,r}^*, \rho_{s,L}^*, \rho_{s,V}^*)
\end{equation}
The subscript $s$ corresponds to segment density, which is used to transform the calculated density as the saturated densities scales significantly number of segments $m_s$ (\textit{cf.} Figure \ref{fig:trans}). Temperature is scaled with respect to the critical temperature ($T^*_r = T^*/T^*_c$) (see Figure \ref{fig:trans} and \ref{fig:trans2}) and vapor pressure is scaled with respect to the critical pressure ($P^*_{v,r} = P^*_v/P^*_c$). For the case of vapor pressure, it is also useful to explore the feasibility of linearizing the temperature-pressure relationship by transforming the reduced temperature and pressure to $1/T_r^*$ and $\ln P_{v,r}^*$ respectively, in agreement with the Clausius–Clapeyron equation\cite{Clausius1850}:
\begin{equation} \label{eq:clausius}
\ln \dfrac{P_2}{P_1} = -\dfrac{\Delta H_v}{R} \left( \dfrac{1}{T_2} - \dfrac{1}{T_1} \right)
\end{equation}
where $\Delta H_v$ is the heat of vaporization and $R$ is the ideal gas constant. Note that the values of $\Delta H_v$ and $R$ are not used in the linearization discussed above.

15,000 VLE data points are generated for 540 different Mie fluids, sampling temperatures between $0.5T_c$ to $0.98 T_c$ at regular intervals of 0.02.

\subsection{Supercritical Density}

In a single supercritical phase, there are two degrees of freedom for state properties, for which temperature $T$ and pressure $P$ are typically set as independent variables. Modelling density of a single phase fluid corresponds to a four-dimensional problem:
\begin{equation}
\mathcal{F}_3 : (m_s, \alpha, T_r^*, P_r^*) \to (\rho_s^*)
\end{equation}
For this case, 27,000 supercritical density data points were generated from 540 different Mie fluids, sampling temperatures between $T_c$ to $2T_c$ and pressures between $P_c$ to $2P_c$, with a uniformly random sampling method. 

\subsection{Model Evaluation}

For each model, the performance and accuracy of the trained machine-learning model is determined by collecting at statistical indicators such as $R^2$ values, mean squared error (MSE) and absolute average deviation (AAD). Although the use of statistical indicators has the advantage of evaluating a model more robustly, it is important to also visually evaluate the individual VLE envelope and isotherms shapes to detect and isolate systematic errors and deviations.

\section{Results}
\subsection{Critical Properties}
\begin{figure}[tbp!]
    \centering
    \includegraphics[scale=1]{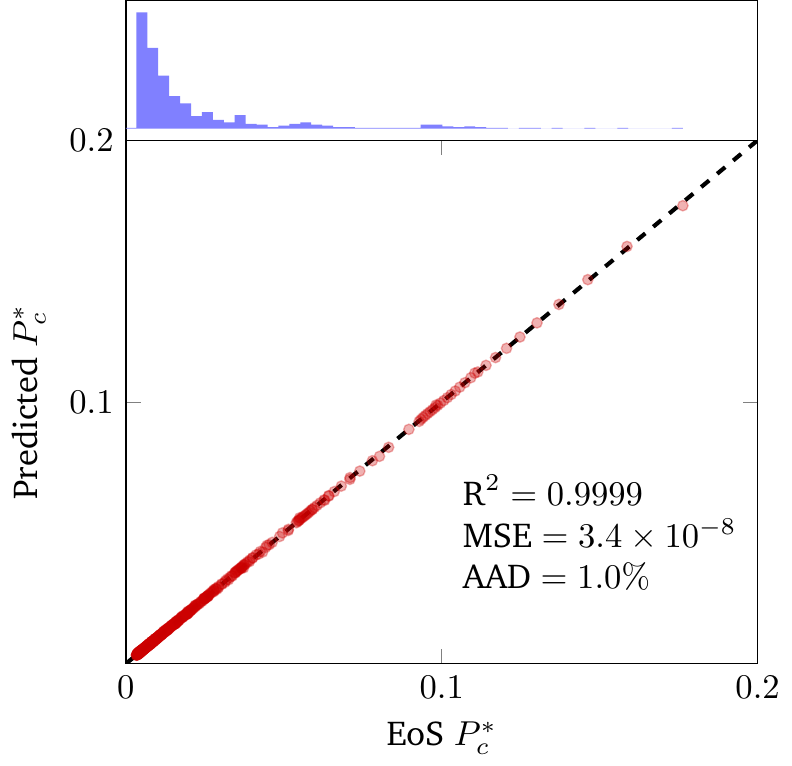}
    \qquad
    \includegraphics[scale=1]{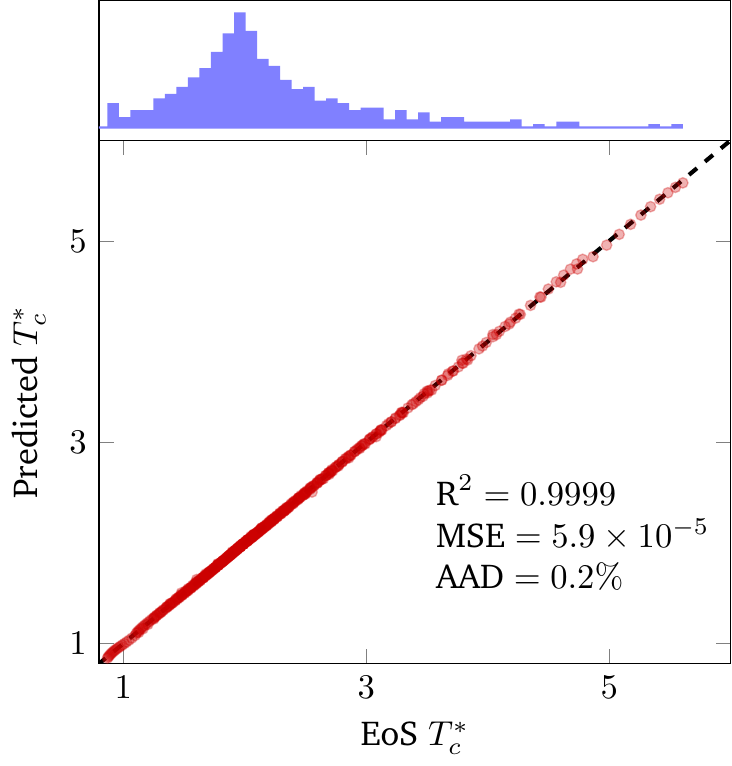}
    \caption{Plot of neural network predicted dimensionless critical pressure $P_c^*$ (left) and critical temperature $T_c^*$ (right) against original SAFT-VR calculated values. Blue histograms describe the distribution of data points.}
    \label{fig:crit-yoyp}
\end{figure}

\begin{table}[htb!]
    \centering
    \begin{tabular}{lll}
        \hline
        \textbf{ANN Specifications} & & \\ \hline
        Input & $m_s$, $\alpha$ & \\
        Output & $P_c^*$, $T_c^*$ & \\
        Hidden Layers & (15, 10, 5) & \\
        Activation Functions & tanh & \\
        Training Data Points & 432 & \\
        Validation Data Points & 108 & \\ \hline
        \textbf{Model Performance} & $P_c^*$ & $T_c^*$ \\ \hline
        Coefficient of Determination $R^2$ & 0.9999 & 0.9999 \\
        Mean Squared Error (MSE) & $3.4 \times 10^{-8}$ & $5.9 \times 10^{-5}$ \\
        Average Absolute Deviation (AAD) & 1.0\% & 0.2\%\\
        \hline
    \end{tabular}
    \caption{Input specifications and model performance for ANN models of critical pressure and critical temperature}
    \label{tab:anncrit}
\end{table}%
Using an ANN with a structure of 3 hidden layers (20,10,5), a ML model was fitted to predict critical pressure and critical temperature between $m_s = 1$ to $20$ and $\lambda_r = 8$ to $34$. With a training data ratio of 0.8, the model performed with high statistical accuracy, achieving $R^2$ values of 0.9999 and absolute average deviation of less than 1\% for both dimensionless critical pressure and critical temperature. Further transformation of the outputs (such as taking the natural log of critical pressure) was not necessary as it did not improve on the performance. Figure \ref{fig:crit-yoyp} shows the scatter plot of all training and validation data points between model predicted values and original data values.

With a combination of RBF and linear kernels, GPR can predict critical points with similar performance ($R^2 = 0.9999$) with only 300 data points. In addition to predicting the critical properties, GPR has an advantage of producing the variances at each predicted point which could be used as an error indicator. In general, this particular feature of GPR is useful as it has a direct application in extensions of this work, as new data points can be included in the EoS tuning by selecting the point with highest variance in the GPR model. 

\begin{figure}[tbp!]
    \centering
    \includegraphics[scale=1]{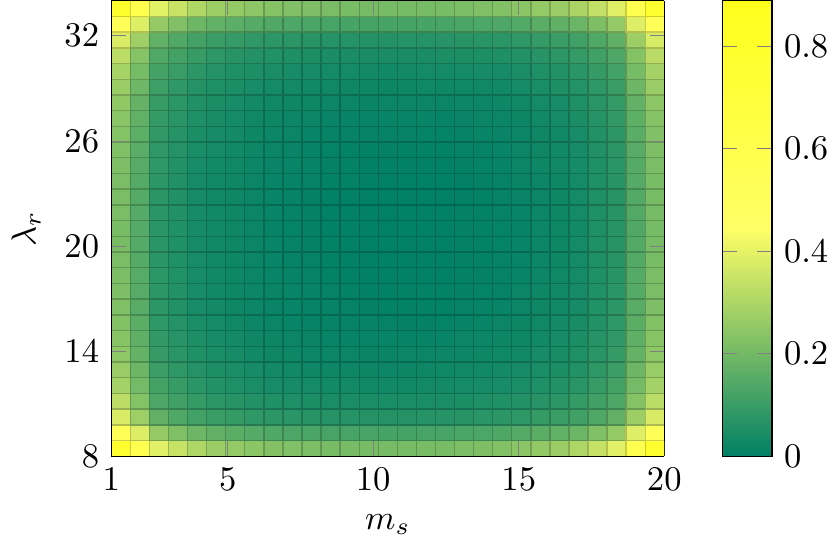}
    \caption{Variance $\sigma^2$ of critical temperature evaluated with Gaussian process regression (GPR) model, as spread across the phase space of molecular descriptors: number of spheres ($m_s$) and repulsive exponent ($\lambda_r$).}
	\label{fig:var}
\end{figure}

In Figure \ref{fig:var}, a colour plot of the variance of critical temperature for the corresponding ranges of $m_s$ and $\lambda_r$ is illustrated. From the GPR model of critical properties, the variance plot shows significant edge effects which causes larger uncertainty for the minimum and maximum values in the range of each inputs. In particular, the corners of the 2-dimensional problem has the largest variance. This, on its own, is not surprising and is an artefact of the way the GPR model fits the data. As a general suggestion, increasing the range beyond the desired working phase space should be considered as a strategy for more accurate predictions.

\subsection{Vapor Pressure}

\begin{table}[htb!]
    \centering
    \begin{tabular}{llll}
        \hline
        \textbf{ANN Specifications} & & & \\ \hline
        Input & $m_s$, $\alpha$, $1/T^*$ &  & \\
        Output & $\ln P_v^*/P_c^*$, $\rho_L^*$, $\rho_V^*$ & & \\
        Hidden Layers & (48, 24, 12) & & \\
        Activation Functions & tanh & & \\
        Training Data Points & 12861 & & \\
        Validation Data Points & 3216 & & \\ \hline
        \textbf{Model Performance} & $P_v^*$ & $\rho_L^*$ & $\rho_V^*$  \\ \hline
        Coefficient of Determination $R^2$ & 0.9985 & 0.9995 & 0.9987 \\
        Mean Squared Error (MSE) & $2.16 \times 10^{-6}$ & $1.37 \times 10^{-5}$ & $5.50 \times 10^{-6}$ \\
        Average Absolute Deviation (AAD) & 4.7\% & 0.5\% & 2.7\% \\
        \hline
    \end{tabular}
    \caption{Input specifications and model performance of ANN models for vapor pressure, saturated liquid density and satuarated vapor density.}
    \label{tab:annvle}
\end{table}%
For vapor pressure, the viability of taking the data transformation based on the Clasius-Clapeyron equation is explored. Using an ANN with a structure of 3 hidden layers (48,24,12), a ML model was fitted to the same range of $m_s$ and $\lambda_r$, with a reduced temperature $T/T_c$ range of 0.5 to 0.95. The predicted output is the vapor pressure reduced with respect to critical pressure ($P_v/P_c$), such that the maximum value is 1.

\begin{figure}[tbp!]
    \centering
    \includegraphics[scale=1]{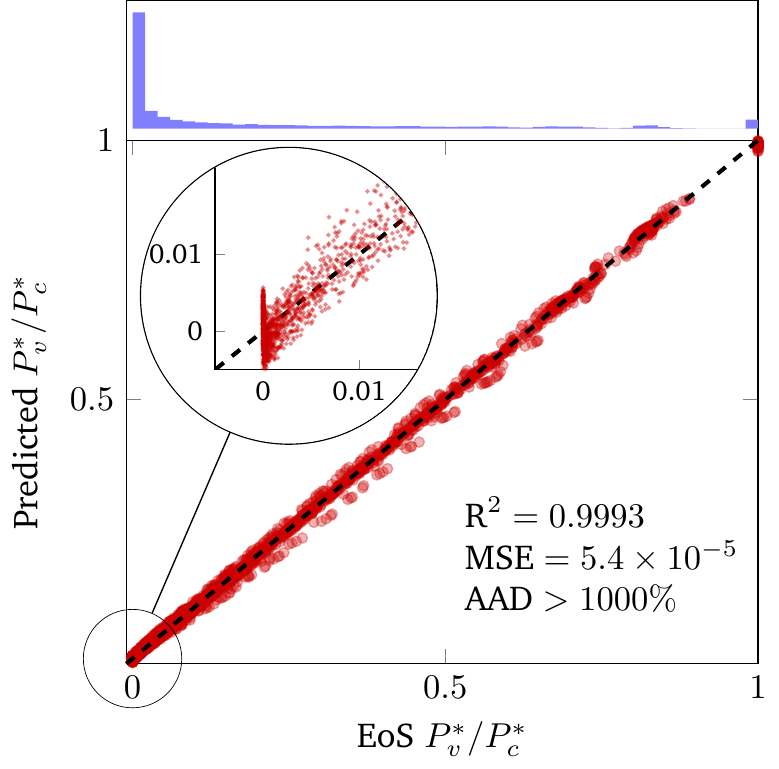}
    \quad
    \includegraphics[scale=1]{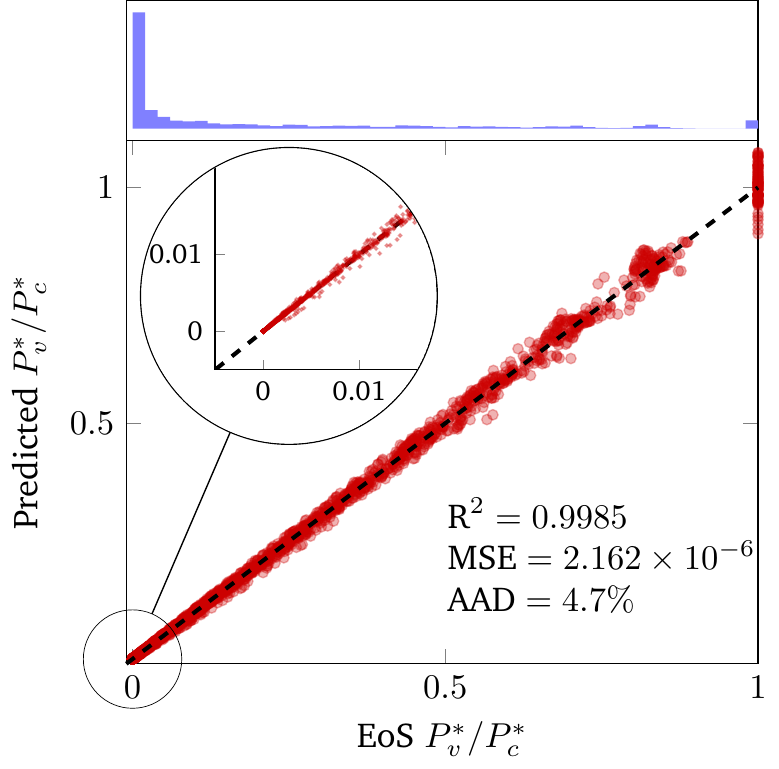}
    \caption{Plot of ANN predicted reduced vapor pressure against benchmark SAFT-VR calculated values, using a non-transformed model with $T^*$ as input and $P^*_v/P^*_c$ as output (left) and a log-transformed model with $1/T^*$ as input and $\ln P^*_v/P^*_c$ as output (right). Blue histograms describe the distribution of data points.}
    \label{fig:pv-yoyp}
\end{figure}

For the normal (unscaled) model, the $R^2$ value of 0.9993 suggests a good model fit for vapor pressure. However, it is important to note that vapor pressure naturally scales logarithmically which results in huge deviations for the very low pressures, resulting in an AAD of over 1000\%. A magnified view of the predicted vs original plot in Figure \ref{fig:pv-yoyp} shows that there are even negative predicted values by the model, which can be interpreted as a small deviation by a statistical analysis but does not make sense from a physical point of view.

\begin{figure}[tbp!]
	\centering
	\includegraphics[scale=1]{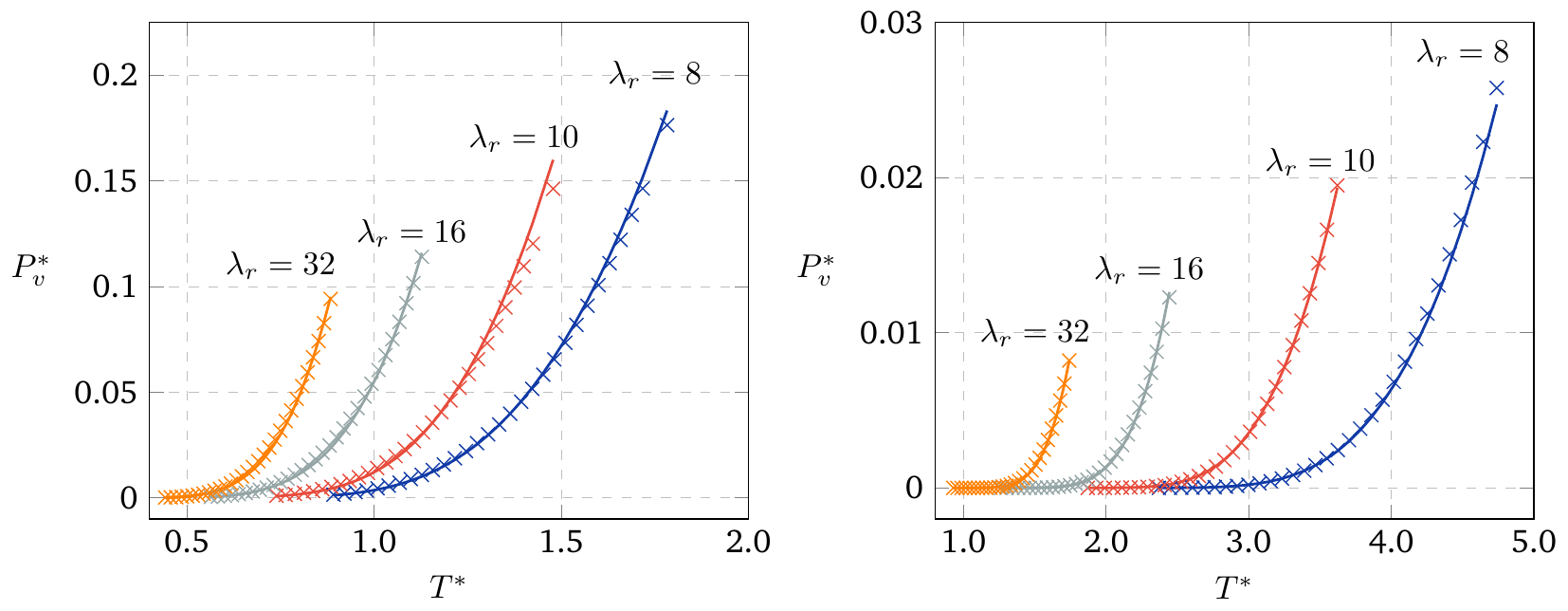}
	\caption{Sample vapor pressure curves predicted by the ANN model (solid line) with data points from the training and validation data set (marks), for $m_s = 1$ (left) and $m_s = 10$ (right).}
	\label{fig:pv-result}
\end{figure}

\begin{figure}[tbp!]
    \centering
    \includegraphics[scale=0.98]{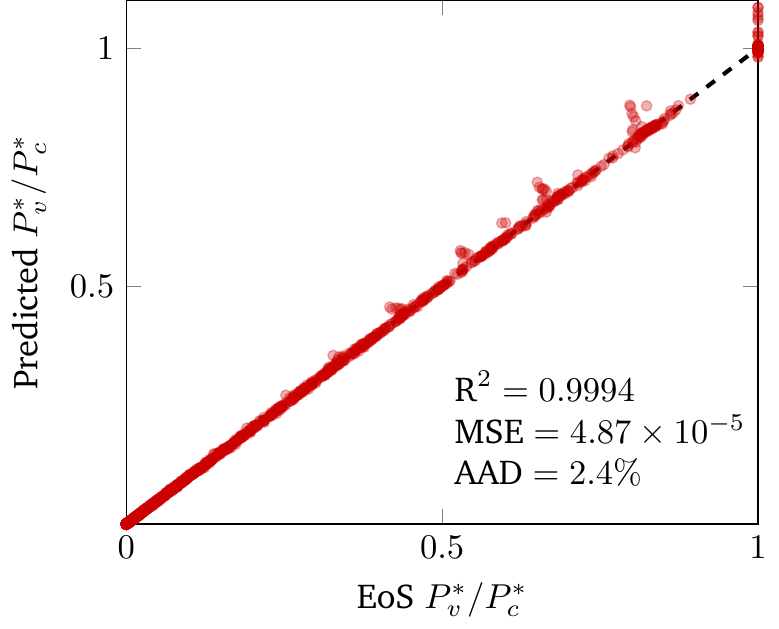}
    \quad
    \includegraphics[scale=0.98]{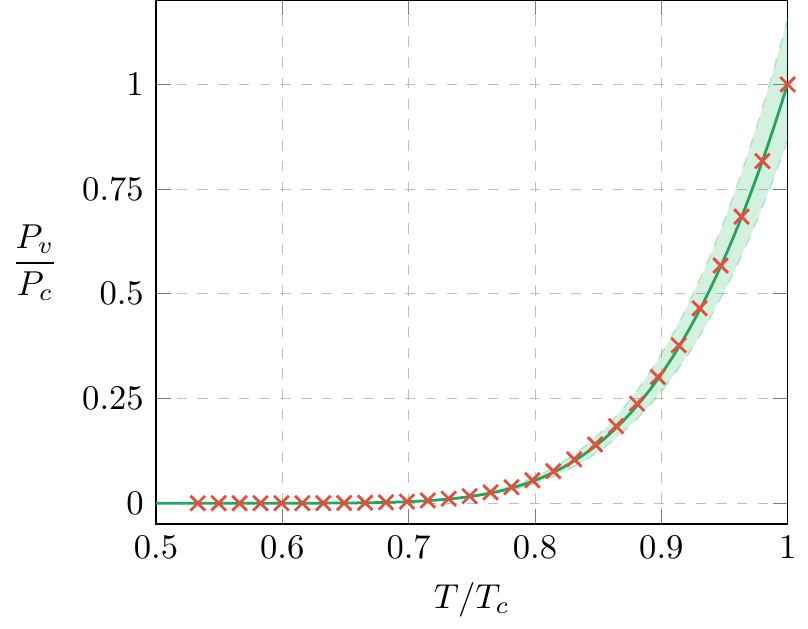}
    \caption{Model performance of GPR model for vapor pressure, with a plot of predicted vapor pressure against original SAFT-VR values (left) and a sample vapor pressure curve with 95\% confidence interval shaded (right) for $m_s = 18$ and $\lambda_r = 14$.}
    \label{fig:pv-gpr-yoyp}
\end{figure}

Taking the Clausius-Clapeyron equation transformation of the P-T data (eq. \ref{eq:clausius}) results in a much better performance for smaller order of magnitude pressures. For the $\ln P_v^*$ and $1/T^*$ model, although deviation at higher values nearer to the critical points are more significant, the overall performance improves with an AAD value of 4.7\%. The magnified view in Figure \ref{fig:pv-yoyp} shows much stronger agreement between the predicted values and the data set at lower orders of magnitude, and inspecting the vapor pressure curves for individual components (Figure \ref{fig:pv-result}) shows that the model managed to capture the shape of the curve successfully.
\begin{table}[htb!]
    \centering
    \begin{tabular}{ll}
        \hline
        \textbf{GPR Specifications} &  \\ \hline
        Input & $m_s$, $\alpha$, $1/T^*$  \\
        Output & $\ln P_v^*/P_c^*$ \\
        Kernel(s) & RBF, Linear \\
        Training Data Points & 320 \\
        Validation Data Points & 15757  \\ \hline
        \textbf{Model Performance} & $P_v^*$ \\ \hline
        Coefficient of Determination $R^2$ & 0.9994 \\
        Mean Squared Error (MSE) & $4.87 \times 10^{-5}$ \\
        Average Absolute Deviation (AAD) & 2.4\% \\
        \hline
    \end{tabular}
    \caption{Input specifications and model performance of a GPR model for vapor pressure.}
    \label{tab:gprpv}
\end{table}%
The downside of taking a natural log is the amplification of error and uncertainty towards larger order of magnitude. This can be seen with a larger spread in Figure \ref{fig:pv-yoyp} (right) where the the inaccuracies increases as the predicted vapor pressure approaches the critical pressure. Using a GPR model taking $m_s$, $\alpha$ and $1/T^*$ as inputs and $\ln (P_v^*/P_c^*)$ as output, for which a good fit is produced with only 320 data points (see Figure \ref{fig:pv-gpr-yoyp} and Table \ref{tab:gprpv}). The shaded area in Figure \ref{fig:pv-gpr-yoyp} (right) shows the 95\% confidence interval for a compound (i.e. $m_s$ and $\lambda_r$ values) unseen by the GPR model, showing the increasing variance (and hence uncertainty) for larger values of $P^*_v/P^*_c$.

\subsection{Saturated Densities}

\begin{figure}[tbp!]
    \centering
    \includegraphics[scale=0.9]{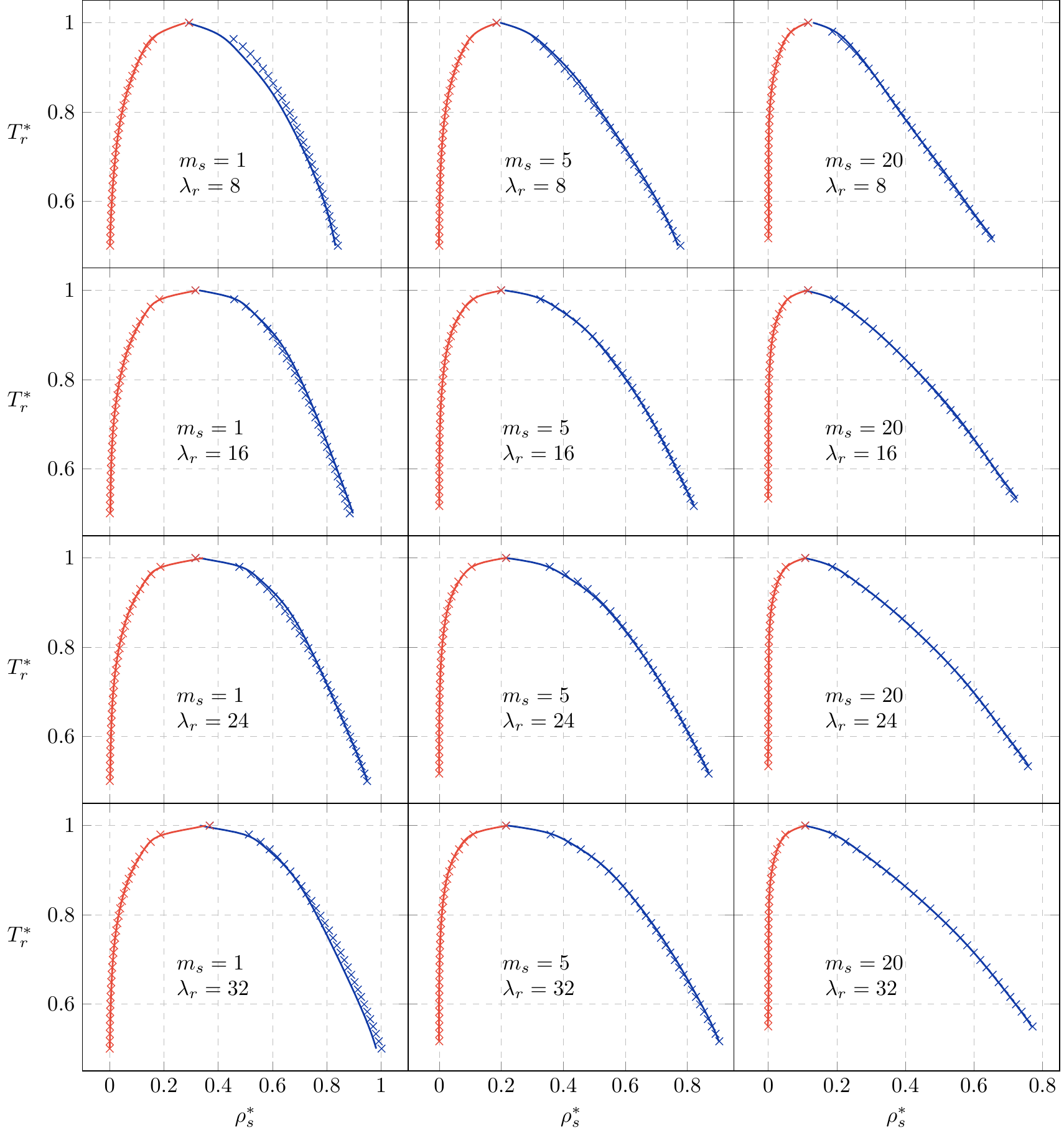}
    \caption{Samples of temperature $T^*_r$ - saturated densities $\rho^*_s$ VLE envelopes for 12 different molecules, corresponding to $m_s = 1, 5, 20$ and $\lambda_r = 8, 16, 24, 32$. Red indicates saturated vapor densities, while blue indicates saturated liquid densities, with solid line representing ANN predicted values, and symbols indicate data points from training and validation data.}
    \label{fig:vle-result}
\end{figure}%
The ANN model employed for vapor pressures, as specified in Table \ref{tab:annvle}, can be used in parallel to predict saturated liquid and vapor densities. We experience that a good model performance as described by common statistical indicators does not necessarily guarantee the correct VLE envelope shape. A visually correct VLE envelope shape was achieved by adding critical temperature and densities into the data set for both saturated liquid and vapor densities. Figure \ref{fig:vle-result} shows the predicted densities next to the benchmark data points for an unbiased selection of 12 different individual molecules. While most molecules shows accurate VLE envelopes, it is clear that the corner of the phase space (e.g. $m_s = 1, \lambda_r = 8$) has visibly less accurate predictions. 

A similar GPR model is developed for saturated densities, which albeit converging and providing acceptable statistical indicators fails to capture the VLE envelope shape even with the inclusion of critical points and employing over 2000 data points in the fitting process. 

\subsection{Supercritical Density}

\begin{table}[htb!]
    \centering
    \begin{tabular}{ll}
        \hline
        \textbf{ANN Specifications} & \\ \hline
        Input & $m_s$, $\alpha$, $T^*$, $P^*$ \\
        Output & $\rho^*$ \\
        Hidden Layers & (48, 24, 12, 6) \\
        Activation Functions & tanh \\
        Training Data Points & 20250 \\
        Validation Data Points & 6750 \\ \hline
        \textbf{Model Performance} & $P_c^*$ \\ \hline
        Coefficient of Determination $R^2$ & 0.9974 \\
        Mean Squared Error (MSE) & $2.55 \times 10^{-3}$ \\
        Average Absolute Deviation (AAD) & 2.4\% \\
        \hline
    \end{tabular}
    \caption{Input specifications and model performance for ANN models of supercritical densities.}
    \label{tab:scden}
\end{table}%
\begin{figure}[tbp!]
    \centering
    \includegraphics[scale=0.98]{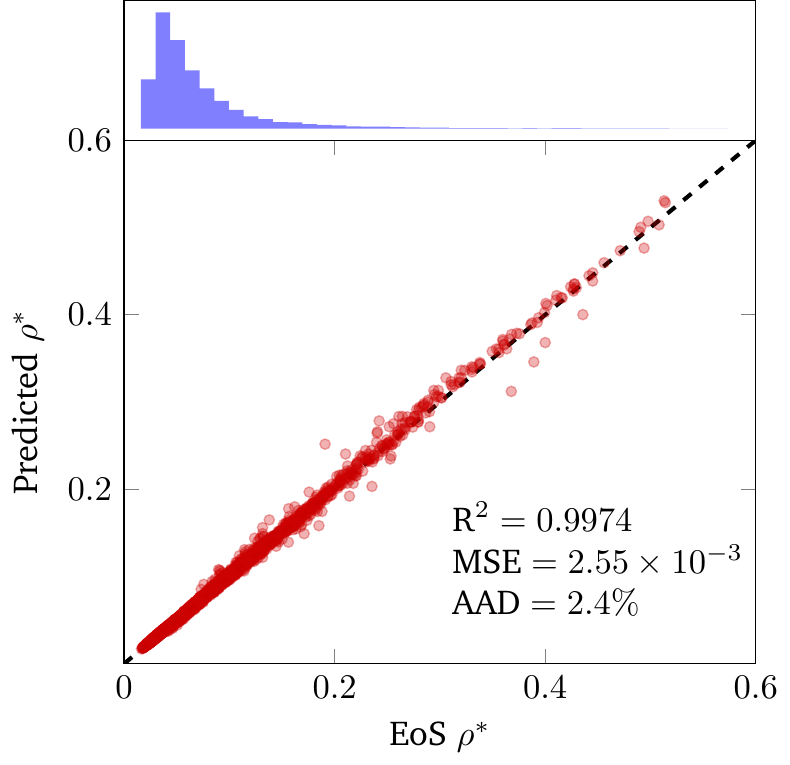}
    \quad
    \includegraphics[scale=0.98]{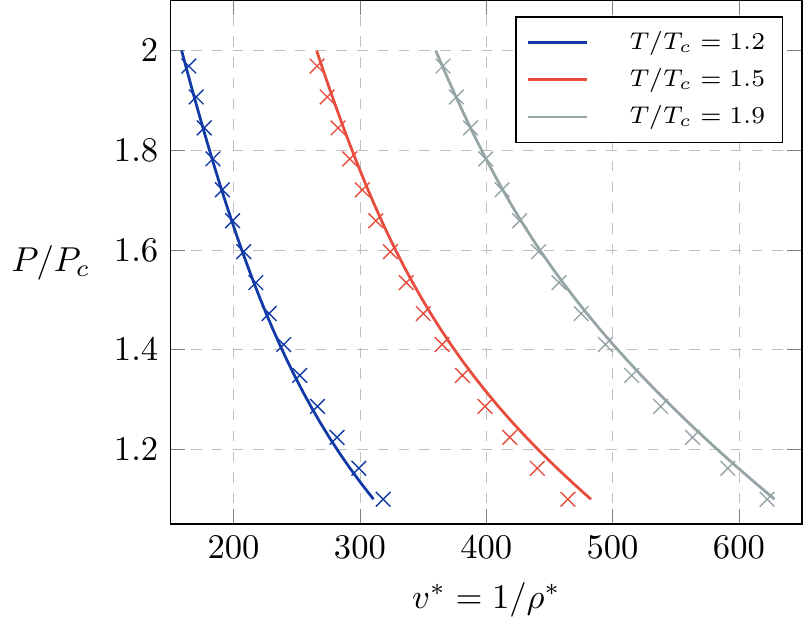}
    \caption{Plot of ANN model predicted supercritical densities against benchmark SAFT-VR calculated values (left), and plot of a sample chain fluid ($m_s = 14$ and $\lambda_r = 21$) with 3 isotherms corresponding to $T/T_c = 1.2, 1.5, 1.9$ (right), with solid line representing ANN model predictions and symbols representing SAFT-VR calculated values for the specific volume ($v^* = 1/\rho^*$) not included in the training set.}
    \label{fig:sc-yoyp}
\end{figure}%
For supercritical densities, the data set was generated using a random sampling method as opposed to at regular interval for VLE properties. The immediate observation is that much more training iterations are required to achieve similar accuracy with ANN. Using an ANN with a structure of 4 hidden layers (48,24,12,6), a model was fitted for the same range of $m_s$ and $\lambda_r$, between reduced temperature $T^*/T^*_c$ and reduced pressure $P^*/P^*_c$ of 1.0 to 2.0. The statistical indicators performed very well even for a 4-input model, with $R^2$ score of 0.9974 and AAD of 2.4\%. Remarkably, randomly selected P-V isotherms within the range of the supercritical temperatures for a random component not included in the training set (Figure \ref{fig:sc-yoyp}) shows that the ANN model managed to capture the general shape of the isotherm. The predicted fluid is a 14-mer chain which, if used to correlate experimental data, would correspond roughly to a C42 alkane (dotetracontane).

\section{Conclusion}

Through the analysis of ML models for different thermodynamic problems, we conclude that both ANN and GPR can be used effectively as surrogates for an analytical EoS. Comparing the two models, GPR requires much less data to achieve a working accuracy and this particular aspect is important if only a reduced experimental data set is available. However, the ability to capture certain shapes of the curves is significantly impeded with the reduced amount of information passed through to GPR, and the computational cost increases significantly with an increase in data set. Fitting through ANNs proves to be a more flexible and robust technique which allows the prediction individual properties with quantitative accuracy and importantly captures the general shapes of different plots of interest. On the downside, ANNs require a much larger data set to train.

The wide range of application of the ML EoS has to be taken into perspective. Although no attempt is made here to match experimental data of real compounds, a value of $m_s = 20$ would roughly correspond to a 60 carbon linear alkane chain (n-hexacontane) and repulsive exponents of $\lambda_r > 20$ are useful to describe highly fluorinated compounds\cite{rahman2018saft}, while a soft potential $\lambda_r = 8$ is essential for modelling water.\cite{lobanova2015saft} The impressive point here is the relative ease with which ML models can both ``develop" and ``learn" an EoS, which in our experience typically requires years of dedicated effort.

There are still many open challenges in the application of ML to predict and correlate thermodynamic data\cite{haghighatlari2019advances,faundez2020misleading}, and the sparsity of real experimental data is a consistent problem. What would happen if computer calculations took over this problem? Modern force fields and algorithms allow the prediction of properties of industrial fluids \textit{in silico} with a level of accuracy comparable to that of experiments with the advantage of being predictive in nature and extensible to regions where experiments would not be practical (e.g. high pressures and temperatures, toxic compounds, etc.). In these cases, gaps in the data can be filled employing classical molecular simulations or in some cases, even quantum mechanical calculations. Existing data can be used to estimate the error (and thus improve on the prediction of the molecular models). More importantly, the vast amount of data (real and pseudo-data) could be “correlated” with ML algorithms. The use of molecular simulation to generate pseudo-experimental data to provide for a sufficiently large data set which can be optimally employed by ML models is an avenue which our group (and others\cite{gong2018predicting,kirch2020brine}) are currently exploring. The product of this enterprise would allow an exponential advance in the efficiency in almost all aspects of bioengineering, energy industries, food and personal care industries, to name but a few.

\begin{acknowledgement}
E.A.M. acknowledges the support from EPSRC through research grants to the Molecular Systems Engineering group (grant nos. EP/E016340, EP/J014958 and EP/R013152). Computations were performed employing the resources of the Imperial College Research Computing Service (DOI: 10.14469/hpc/2232) and the UK Materials and Molecular Modelling Hub, which is partially funded by EPSRC (grant no. EP/P020194).
\end{acknowledgement}


\bibliography{ref}

\end{document}